\documentclass[namedreferences]{solarphysics}

\usepackage[hyperref,optionalrh,showbiblabels]{spr-sola-addons} 
\usepackage{graphicx}        

\usepackage{xspace}

\renewcommand{\vec}[1]{{\mathbfit #1}}

\chardef\us=`\_

\newcommand{\omits}[1]{}

\begin{document}

\begin{article}
\begin{opening}

\title{Inferring flare loop parameters with measurements of standing sausage modes}

\author[addressref={aff1,aff2}]{\inits{}\fnm{Ming-Zhe}~\lnm{Guo}}
\author[addressref={aff1}]{\inits{}\fnm{Shao-Xia}~\lnm{Chen}}
\author[addressref={aff1},corref,email={bbl@sdu.edu.cn}]
{\inits{Bo}\fnm{Bo}~\lnm{Li}}
\author[addressref={aff1}]{\inits{}\fnm{Li-Dong}~\lnm{Xia}}
\author[addressref={aff1}]{\inits{}\fnm{Hui}~\lnm{Yu}}

\address[id=aff1]{Shandong Provincial Key Laboratory of Optical Astronomy and
Solar-Terrestrial Environment, Institute of Space Sciences, Shandong University, Weihai, 264209, China}

\address[id=aff2]{CAS Key Laboratory of Geospace Environment, University of Science \& Technology of China, Chinese Academy of Sciences, Hefei 230026, China}

\runningauthor{Guo et al.}
\runningtitle{Inferring flare loops parameters with sausage mode measurements}

\begin{abstract}
Standing fast sausage modes in flare loops 
    were suggested to account for
    a considerable number of quasi-periodic pulsations (QPPs) in the light curves of solar flares.
This study continues our investigation into 
    the possibility to invert the measured periods $P$ and damping times $\tau$  
    of sausage modes
    to deduce the transverse Alfv\'en time $R/v_{\rm Ai}$,
    density contrast $\rho_{\rm i}/\rho_{\rm e}$,
    and the steepness of the density distribution transverse to flare loops.
A generic dispersion relation (DR) governing linear sausage modes
    is derived for pressureless cylinders where density inhomogeneity of arbitrary form
    takes place within the cylinder.
We show that in general the inversion problem is under-determined for QPP events where only a single
    sausage mode exists, be the measurements spatially resolved or unresolved.
While $R/v_{\rm Ai}$ can be inferred to some extent, the range of possible steepness parameters 
    may be too broad to be useful.
However, for spatially resolved measurements where an additional mode is present,
    it is possible to deduce self-consistently $\rho_{\rm i}/\rho_{\rm e}$, the profile steepness,
    and the internal Alfv\'en speed $v_{\rm Ai}$.
We show that at least for a recent QPP event that involves a fundamental kink mode in addition to a sausage one,
    flare loop parameters are well constrained, even if the specific form
    of the transverse density distribution remains unknown.
We conclude that spatially resolved, multi-mode QPP measurements need to be pursued for inferring
    flare loop parameters. 
\end{abstract}
\end{opening}

\section{Introduction}
\label{sec_intro}
There is now ample evidence for the existence of low-frequency waves and oscillations
     in the structured solar atmosphere
     \citep[for recent reviews, see e.g.,][]{2005LRSP....2....3N,2007SoPh..246....3B,2008IAUS..247....3R,2012RSPTA.370.3193D,2014SoPh..289.3233L}.
When combined with magnetohydrodynamic (MHD) theory, 
     the measured wave parameters allow one
     to infer the solar atmospheric parameters that are difficult to measure directly.
This practice was originally proposed for the solar corona
     (\citeauthor{1984ApJ...279..857R}~\citeyear{1984ApJ...279..857R},
     see also \citeauthor{1970A&A.....9..159R}~\citeyear{1970A&A.....9..159R},
     \citeauthor{1970PASJ...22..341U}~\citeyear{1970PASJ...22..341U},
     \citeauthor{1975IGAFS..37....3Z}~\citeyear{1975IGAFS..37....3Z}),
     but has been extended to spicules~\citep[e.g.,][]{2009SSRv..149..355Z},
     prominences~\citep[e.g.,][]{2012LRSP....9....2A},
     magnetic pores~\citep[e.g.,][]{2011ApJ...729L..18M},
     and various structures in the chromosphere~\citep[e.g.,][]{2009Sci...323.1582J,2012NatCo...3E1315M},
     to name but a few.
Compared with sausage waves (with azimuthal wavenumber $m=0$),
     kink waves (with $m=1$) have received more attention,
     presumably due to their ubiquity in the solar atmosphere
     \citep[e.g.,][]{1999Sci...285..862N,1999ApJ...520..880A,2009ApJ...697.1384T,2013SoPh..284..559K}.
However, recent observations indicated that sausage waves abound
     as well \citep[e.g.,][]{2003A&A...412L...7N,2012NatCo...3E1315M,2015ApJ...806..132G,2015A&A...578A..60M}.
In addition, standing sausage modes in flare loops 
     were shown to be 
     important for interpreting a considerable fraction of
     quasi-periodic pulsations (QPPs) in the lightcurves of solar flares
     \citep[see][for a recent review]{2009SSRv..149..119N}. 

A theoretical understanding of fast sausage waves supported by magnetized cylinders
    is crucial for their seismological applications.
For this purpose, the transverse density distribution
    is usually idealized as being in a step-function (top-hat) form, characterized by
    the internal ($\rho_{\rm i}$) and external ($\rho_{\rm e}$) values
    \citep[e.g.,][]{1978SoPh...58..165M, 1982SoPh...75....3S, 1983SoPh...88..179E,1986SoPh..103..277C,2007AstL...33..706K,
    2014ApJ...781...92V}.
In a low-$\beta$ environment such as the solar corona, two regimes
    are known to exist, depending on the longitudinal wavenumber $k$
    \citep[e.g.,][]{1982SoPh...75....3S}.
When $k$ exceeds some critical $k_{\rm c}$,
    the trapped regime arises whereby the sausage wave energy is well confined to
    the cylinder.
On the contrary, if $k< k_{\rm c}$, then the leaky regime results and fast sausage waves
    experience apparent temporal damping by emitting their energy into
    the surrounding fluid.
Furthermore, the $k$-dependence of the periods $P$ and damping times $\tau$ of leaky waves
    disappears when $k$ is sufficiently small~\citep[e.g.,][]{2007AstL...33..706K, 2014ApJ...781...92V}.
Let $R$ denote the cylinder radius, 
    and $v_{\rm Ai}$ denote the internal Alfv\'en speed.
In the long-wavelength limit ($k\rightarrow 0$),
    $P$ is found to be primarily determined
    by the transverse Alfv\'en transit time $R/v_{\rm Ai}$, while
    the ratio $\tau/P$ is largely proportional to the density contrast $\rho_{\rm i}/\rho_{\rm e}$
    \citep{2007AstL...33..706K}.
This then enables one to employ the measured $P$ and $\tau$ 
    to deduce $\rho_{\rm i}/\rho_{\rm e}$ and $R/v_{\rm Ai}$, with the latter carrying
    important information on the magnetic field strength in flare loops.

Evidently, there is no reason to expect that the density distribution across magnetic cylinders
    is in a step-function fashion.
This has stimulated a series of studies to examine the properties of fast sausage waves
    in magnetized cylinders with a continuous transverse density profile
    by either proceeding analytically with an 
    {eigen-mode analysis
    \citep{1988A&A...192..343E,2014A&A...572A..60L,2015ApJ...810...87L}}
    or numerically solving the linearized MHD equations as an initial-value problem
    \citep{2012ApJ...761..134N,2015SoPh..290.2231C}.
Many features in the step-function case, the $k$-dependence in particular, were found to survive.
However, the period $P$~\citep{2012ApJ...761..134N} and damping time $\tau$~\citep{2015SoPh..290.2231C}
    may be sensitive to yet another parameter,
    namely the steepness or equivalently the lengthscale of the transverse density inhomogeneity.
Note that the steepness is crucial in determining such coronal heating mechanisms
    as resonant absorption \citep[e.g.,][]{1988JGR....93.5423H,2002A&A...394L..39G,2002ApJ...577..475R}
    and phase mixing \citep{1983A&A...117..220H}.
There is then an obvious need to employ the measured $P$ and $\tau$ of sausage modes
    to infer the profile steepness,
    in much the same way that kink modes were employed \citep{2007A&A...463..333A,2008A&A...484..851G,2014ApJ...781..111S}.
This was undertaken by \citeauthor{2015ApJ...812...22C} (\citeyear{2015ApJ...812...22C}, hereafter paper I),
    based on an analytical dispersion relation (DR) governing linear fast sausage waves
    in cylinders with a rather general transverse density distribution.
The only requirement was that this profile can be decomposed into a uniform cord,
    a uniform external medium, and a transition layer connecting the two.
However, this layer can be of arbitrary width and the profile therein can be
    in arbitrary form.

The aim of the present study is to extend the analysis in paper I in {the following aspects}.
First, we will remove the restriction for the transverse density profile
    to involve a uniform cord, thereby enabling the analysis to be applicable to
    a richer variety of density distributions.
{Second, when validating the results from this eigenmode analysis, we employ an independent approach
    by solving the time-dependent version of linear MHD equations.
    We will detail these time-dependent computations pertinent to the afore-mentioned transverse density profile.
    }    
{Third}, we will extend the seismological applications in paper I to QPP events
    that involve both kink and sausage modes.
To illustrate the scheme for inverting multi-mode measurements,
    paper I adopted the analytical expressions for the kink mode period
    and damping time in the thin-tube-thin-boundary (TTTB) limit
    as given by \cite{2008A&A...484..851G}.
In this study we replace the TTTB expressions with a self-consistent, linear, resistive MHD computation.
This is necessary given that flare loops tend not to be thin \citep{2004ApJ...600..458A},
    and it is not safe to assume a priori that the density inhomogeneity
    takes place in a thin transition layer.
{Fourth, in connection with the third point, we take this opportunity to provide a rather detailed examination of 
    resonantly damped kink modes in cylinders with transverse density profiles in question.
}    

This manuscript is organized as follows.
Section~\ref{sec_model} presents the necessary equations, the derivation
    of the analytic DR in particular.
The behavior of sausage waves in nonuniform cylinders and
    its applications to QPP events
    are then presented in Sect.~\ref{sec_num_res}.
Finally, Sect.~\ref{sec_con} summarizes the present study.

\section{Mathematical Formulation}
\label{sec_model}
\subsection{Derivation of the Dispersion Relation}
Appropriate for the solar corona, we adopt ideal, cold (zero-$\beta$) MHD to describe 
     fast sausage waves.
The magnetic loops hosting these waves are modeled as straight cylinders
     with radius $R$ aligned with a uniform magnetic field $\vec{B} = B\hat{z}$, where
     a standard cylindrical coordinate system $(r, \theta, z)$ is adopted.
The equilibrium density is assumed to be a function of $r$ only and of the form
\begin{equation}
\label{eq_density}
\rho(r)=\left\{
\begin{array}{cc}
\rho_{\rm i}\left[1-\left(1-\displaystyle\frac{\rho_{\rm e}}{\rho_{\rm i}}\right)f(r)\right],   & 0 \le r < R,\\
\rho_{\rm e}, & r > R,
\end{array}\right.
\end{equation}
    where $f(r)$ is some arbitrary function that increases smoothly from $0$ at $r=0$
    to unity when $r=R$.
Furthermore,   
    $\rho_{\rm i}$ and $\rho_{\rm e}$ denote the densities at the cylinder axis
    and in the external medium, respectively. 
{The corresponding Alfv\'en speeds follow from the definition
    $v_{\rm Ai, e} = B/\sqrt{4\pi\rho_{\rm i, e}}$.
}    
    
{ 
It suffices to briefly outline the mathematical approach for establishing the pertinent dispersion relation (DR),
   since this approach has been detailed in paper I. 
To start, we specialize to axisymmetric sausage perturbations, and Fourier-analyze any perturbation $\delta f(r, z, t)$ as
\begin{eqnarray}
  \delta f(r, z, t) = {\rm Re}\left\{\tilde{f}(r)\exp\left[-i\left(\omega t-k z\right)\right]\right\} . 
  \label{eq_Fourier_Ansatz}
\end{eqnarray}
It then follows from the linearized, ideal, cold MHD equations that the 
    Fourier amplitudes of the transverse Lagrangian displacement ($\tilde{\xi}_r$)
    and Eulerian perturbation of total pressure ($\tilde{p}_{\rm T}$) are 
    governed by Eqs.~(6) and (7) in paper I, respectively.
Now that sausage waves do not resonantly couple to torsional Alfv\'en waves for the configuration we examine,
    one may employ regular series expansions about $y\equiv r-R/2 = 0$ to express $\tilde{\xi}_r$ and $\tilde{p}_{\rm T}$
    in the nonuniform portion of the density distribution.
Further requiring that sausage waves do not disturb the cylinder axis ($\tilde{\xi}_r = 0$ at $r=0$),
    and employing the conditions for $\tilde{\xi}_r$ and $\tilde{p}_{\rm T}$ to be continuous at the interface $r=R$,
    one finds that the DR can be expressed as
\begin{equation}
\label{eq_DR}
\begin{array}{rcl}
 && \displaystyle\frac{\displaystyle\frac{\mu_{\rm e}R H_0^{(1)}(\mu_{\rm e} R)}{H_1^{(1)}(\mu_{\rm e} R)}\tilde{\xi}_{1}(R/2)-\tilde{\xi}_{1}(R/2)-R\tilde{\xi}^\prime_{1}(R/2)}{\tilde{\xi}_{1}(-R/2)} \\ [0.5cm] &=&\displaystyle\frac{\displaystyle\frac{\mu_{\rm e}R H_0^{(1)}(\mu_{\rm e} R)}{H_1^{(1)}(\mu_{\rm e} R)}\tilde{\xi}_{2}(R/2)-\tilde{\xi}_{2}(R/2)-R\tilde{\xi}^\prime_{2}(R/2)}{\tilde{\xi}_{2}(-R/2)}~.
 \end{array}
\end{equation}
Here $H_n^{(1)}$ denotes the $n$th-order Hankel function of the first kind,
    and $\mu_{\rm e}$ is defined by
\begin{eqnarray}
 \label{eq_def_mue}
    \mu_{\rm e}^2 = {\frac{\omega^2}{v^2_{\rm Ae}}-k^2}~~~~~~~
    (-\frac{\pi}{2} < \arg \mu_{\rm e} \le \frac{\pi}{2}) .
 \end{eqnarray}
Furthermore,
\begin{equation}\label{eq_two}
   \tilde{\xi}_{1}(y)=\sum^\infty_{n=0}a_n y^{n}~~\mbox{and}~~
   \tilde{\xi}_{2}(y)=\sum^\infty_{n=0}b_n y^{n}
\end{equation}
    are two linearly independent solutions for $\tilde{\xi}_r$ in the portion $r<R$.
Without loss of generality, we choose
\begin{eqnarray}
\label{eq_a01_b01}
   a_0=R,~~a_1=0,~~b_0=0,~~b_1 =1 .
\end{eqnarray}
The rest of the coefficients $a_n$ and $b_n$ can be found by replacing $R$ with $R/2$ in Eq.~(11) in paper I,
    and contain the information on the density distribution.
Finally, the prime $'$ denotes the derivative of $\tilde{\xi}_{1, 2}$ with respective to $y$.

Before proceeding, we note that a series-expansion-based approach was recently adopted by \citet{2013ApJ...777..158S}
    to treat wave modes in transversally nonuniform cylinders where the azimuthal wavenumber $m$
    is allowed to be arbitrary.
A comparison between that approach and ours is detailed in Appendix~\ref{sec_App_compareSeries},   
    where we show that both approaches yield identical results for trapped sausage modes ($m=0$).
While our approach seems more appropriate to describe leaky sausage modes,
    we stress that a singular expansion as employed by \citet{2013ApJ...777..158S}
    is necessary to treat wave modes with $m\ne 0$.
    }

\subsection{Method of solution}
Throughout this study, we focus on standing sausage modes by restricting longitudinal
    wavenumbers ($k$) to be real, but allowing angular frequencies ($\omega$)
    to be complex-valued ($\omega = \omega_{\rm R}+i\omega_{\rm I}$).
In addition, we focus on fundamental modes, namely those with $k=\pi/L$ where
    $L$ is the loop length.
In practice, we start with prescribing an $f(r)$, and then solve Eq.~(\ref{eq_DR})
    by truncating the infinite series expansion (Eq.~\ref{eq_two})
    to retain terms with $n$ up to $N=101$.
Using an even larger $N$ leads to no appreciable difference.
It should be noted that $\omega$ in units of $v_{\rm Ai}/R$
    depends only on the combination $[f(r),~kR,~\rho_{\rm i}/\rho_{\rm e}]$.
The corresponding period $P$
    and damping time $\tau$ follow from the definitions $P=2\pi/\omega_{\rm R}$
    and $\tau = 1/|\omega_{\rm I}|$.

For validation purposes, we also obtain $\omega$ as a function of $k$ in a way
    independent from this eigen-mode analysis.
This is done by solving the time-dependent equation governing the transverse velocity perturbation $\delta v_r (r, z,t) $
    as an initial-value problem.
{
For given combinations of $[f(r), kR, \rho_{\rm i}/\rho_{\rm e}]$,
   one may find the periods and damping times of sausage modes
   by analyzing the temporal evolution of the perturbation signals (see Appendix~\ref{sec_App_tdependent} for details).
}    
As will be shown by Fig.~\ref{fig_Ptau_vs_L}, the values of $P$ and $\tau$
   derived from the two independent approaches are in close agreement.
However, numerically solving the analytical DR is much less computationally expensive.
On top of that, the values of $\tau$ for heavily damped modes can be readily found,
   whereas the perturbation signals in time-dependent computations
   decay too rapidly to allow a proper determination of $\tau$.

\section{Numerical Results}
\label{sec_num_res}

\begin{figure}
\centerline{\includegraphics[width=0.8\columnwidth]{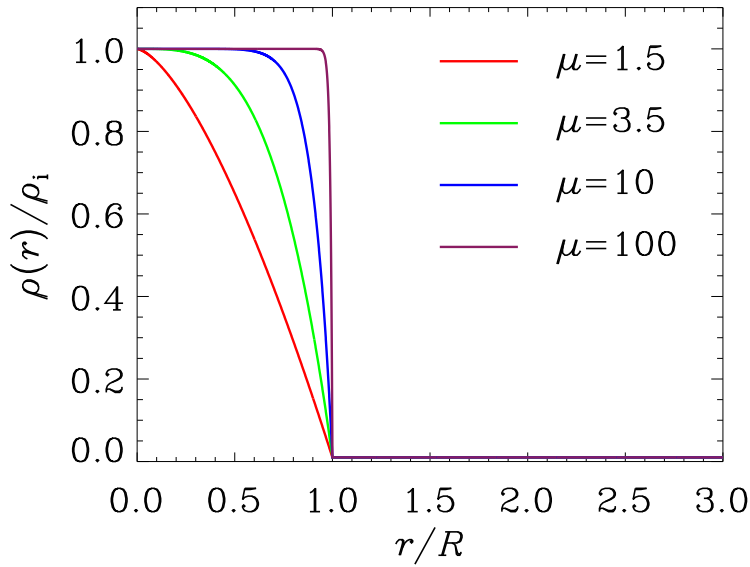}}
 \caption{
 Transverse equilibrium density profiles as a function of $r$
    for different steepness parameters $\mu$ as labeled.
 Here the density contrast $\rho_{\rm i}/\rho_{\rm e}$
    is chosen to be $100$ for illustration purposes.
}
 \label{fig_illus_profile}
\end{figure}

It is impossible to exhaust the possible prescriptions for $f(r)$.
We therefore focus on one choice, namely
\begin{eqnarray}
\label{eq_def_fr}
  f(r)= \left(\frac{r}{R}\right)^\mu ,
\end{eqnarray}
    where $\mu$ is positive.
The density profiles with a number of different $\mu$ are shown
    in Fig.~\ref{fig_illus_profile}, where
    $\rho_{\rm i}/\rho_{\rm e}$ is chosen to be $100$
    for illustration purposes.
Evidently,
    the profile becomes increasingly steep as $\mu$ increases
    and approaches a step-function form when $\mu$ approaches infinity.
This makes it possible to investigate the effect of profile steepness by
    examining the $\mu$-dependence of the numerical results.
In addition, for fundamental modes with $k=\pi/L$, the dependence     
    on $kR$ is translated into that on the length-to-radius ratio $L/R$.

\subsection{Behavior of sausage waves in nonuniform tubes}

\begin{figure}
\centerline{\includegraphics[width=0.8\columnwidth]{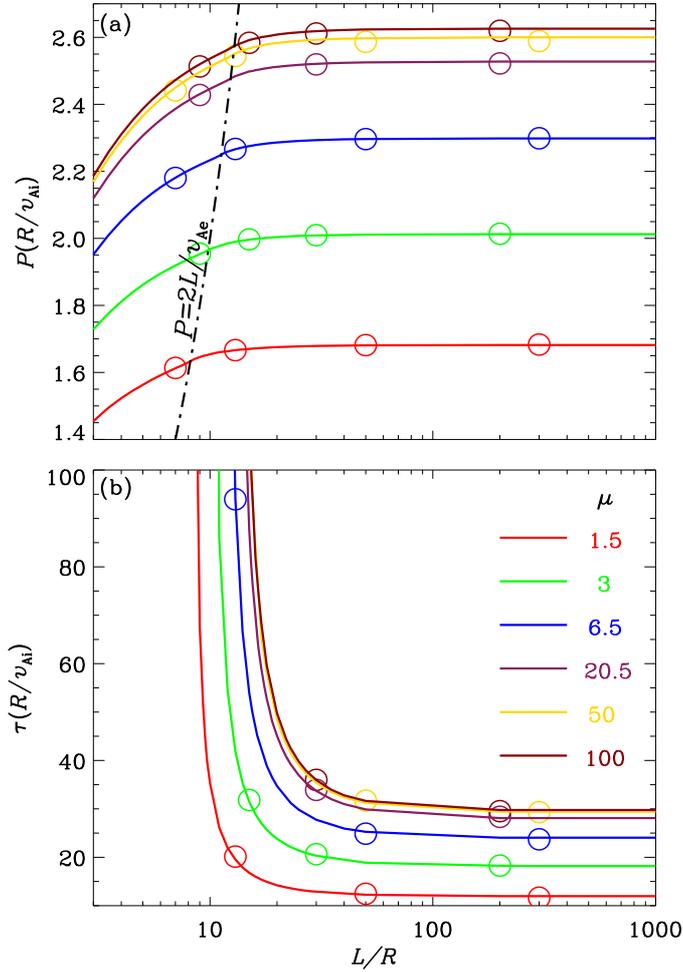}}
 \caption{
 Dependence on length-to-radius ratio $L/R$ of (a) periods $P$ and
      (b) damping times $\tau$ of fundamental sausage modes.
 A number of density profiles with different $\mu$ are examined as labeled.
 The black dash-dotted line in (a) represents $P=2L/v_{\rm Ae}$ and separates
      the trapped (to its left) from leaky (right) regimes.
 The open circles represent the values obtained by solving Eq.~(\ref{eq_xi(r,t)})
      with an initial-value-problem approach, which
      is independent from the eigen-mode analysis presented in the text.
 The density contrast $\rho_{\rm i}/\rho_{\rm e}$ is chosen to be $100$.
 }
 \label{fig_Ptau_vs_L}
\end{figure}

Figure~\ref{fig_Ptau_vs_L} presents the dependence on $L/R$ of the period $P$
    and damping time $\tau$ for a series of $\mu$ values
    as labeled.
For illustration purposes, the density ratio
    $\rho_{\rm i}/\rho_{\rm e}$ is taken to be $100$.
The dash-dotted line in Fig.~\ref{fig_Ptau_vs_L}a represents
    $P=2L/v_{\rm Ae}$, and separates trapped (to its left)
    from leaky (right) modes.
The solid curves represent the results from solving the analytical DR (Eq.~\ref{eq_DR}),
    whereas the circles represent those obtained with the initial-value-problem approach.
A close agreement between the curves and circles is clear,
    thereby validating the DR.

Figure~\ref{fig_Ptau_vs_L}a indicates that
    the wave period $P$ increases monotonically with $L/R$ in the trapped regime,
    and rapidly settles to some asymptotic value in the leaky one.
Likewise, Fig.~\ref{fig_Ptau_vs_L}b shows that, being
    identically infinite in the trapped regime,
    the damping time $\tau$
    also experiences saturation for sufficiently large $L/R$.
In addition, both $P$ and $\tau$ 
    increase substantially
    with increasing $\mu$ at a given $L/R$.
We note that while the tendency for $P$ to be larger for steeper density profiles
    agrees with the study by \citet{2012ApJ...761..134N},
    it does not hold in general.
As a matter of fact, Fig.~3 in paper I shows that the opposite occurs
    for some different profile prescriptions.
This means that the largely unknown specific form of the transverse density distribution
    plays an important role in determining the dispersive properties of sausage modes.
Consequently, when the period and damping time of sausage modes are seismologically exploited,
    the uncertainty in specifying the density profile needs
    to be considered.

\subsection{Applications to spatially unresolved QPP observations}
In essence, Fig.~\ref{fig_Ptau_vs_L} indicates that the period $P$ and damping time $\tau$ of sausage modes
    can be formally expressed as
\begin{eqnarray}
&& P_{\rm saus}
    =\displaystyle\frac{R}{v_{\rm Ai}}F_{\rm saus}\left(\frac{L}{R},~\mu,~
    \frac{\rho_{\rm i}}{\rho_{\rm e}}\right),
    \label{eq_Psaus} \\
&& \displaystyle\frac{\tau_{\rm saus}}{P_{\rm saus}}
    = G_{\rm saus}\left(\frac{L}{R},~\mu,~\frac{\rho_{\rm i}}{\rho_{\rm e}}\right) .
    \label{eq_tauOPsaus}
\end{eqnarray}
Note that the damping-time-to-period ratio $\tau/P$ is adopted here instead of $\tau$ itself,
    the reason being that $\tau/P$ does not depend on $R/v_{\rm Ai}$.
Furthermore, the $L/R$-dependence
    disappears for cylinders with large enough $L/R$.

Let us first consider the applications of Eqs.~(\ref{eq_Psaus}) and (\ref{eq_tauOPsaus}) to spatially unresolved QPP events, for which
    only $P$ and/or $\tau$ can be regarded known.
However, the information is missing on both the physical parameters $[v_{\rm Ai},~\mu,~\rho_{\rm i}/\rho_{\rm e}]$
    and geometrical parameters $[L,~R]$.
If a trapped sausage mode is responsible for causing a QPP event as happens
    when the signals do not show clear damping,
    then only Eq.~(\ref{eq_Psaus}) is relevant.
This means that any point on a 3-dimensional (3D) hypersurface in the 4D space formed by
    $[R/v_{\rm Ai},~L/R,~\mu,~\rho_{\rm i}/\rho_{\rm e}]$ is possible to reproduce the measured $P$.
Even if the signals in a QPP event are temporally decaying,
    the range of possible parameters that can reproduce the measured $P$ and $\tau$
    is still too broad to be useful: 
    a 2D surface in the 4D parameter space results.
The situation improves if one can assume that the flare loops
    hosting sausage modes are sufficiently thin
    such that the $L/R$-dependence drops out.
Equations~(\ref{eq_Psaus}) and (\ref{eq_tauOPsaus}) then suggest that for trapped (leaky) modes, one can deduce
    a 2D surface (1D curve) in the 3D space formed by $[R/v_{\rm Ai},~\mu,~\rho_{\rm i}/\rho_{\rm e}]$.
{We note that the idea for deriving 1D inversion curves was first introduced by 
    \citet{2007A&A...463..333A} and later explored 
    in e.g., \citet{2008A&A...484..851G,2014ApJ...781..111S}.
While resonantly damped kink modes were examined therein,
    the same idea applies also to leaky sausage modes in thin cylinders,
    the only difference being that the transverse Alfv\'en time ($R/v_{\rm Ai}$)
    replaces the longitudinal one ($L/v_{\rm Ai}$). 
}  

\begin{figure}
\centering
\includegraphics[width=0.6\columnwidth]{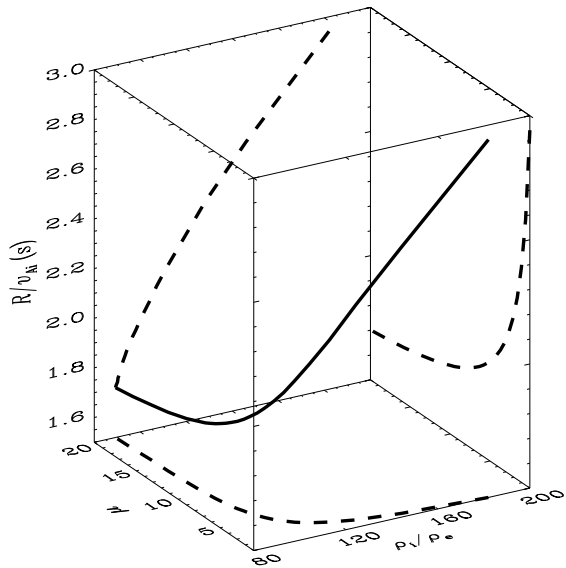}
 \caption{
  Inversion curve (the solid line) and its projections (dashed)
      in the three-dimensional parameter space
      spanned by $[\mu,~\rho_{\rm i}/\rho_{\rm e},~R/v_{\rm Ai}]$.
  All points along this curve are equally possible to reproduce
      the quasi-periodic-pulsation event reported in~\citet{1973SoPh...32..485M}, where
      the oscillation period is $4.3$~secs,
      and the damping-time-to-period ratio is $10$.}
\label{fig_3D_results}
\end{figure}

Figure~\ref{fig_3D_results} presents the 1D curve and its projections (the dashed lines)
    onto various planes in the
    $[R/v_{\rm Ai},~\mu,~\rho_{\rm i}/\rho_{\rm e}]$ space,
    using the QPP event reported in \citet{1973SoPh...32..485M} as an example.
For this event, the oscillation period is $4.3$~secs,
      and the damping-time-to-period ratio is $10$.
Table~\ref{tab_invcurv} presents a set of values read from the solid curve in Fig.~\ref{fig_3D_results}.
One can see that among the parameters to infer,
    the transverse Alfv\'en time $R/v_{\rm Ai}$ and
    density ratio $\rho_{\rm i}/\rho_{\rm e}$ can be somehow constrained.
To be specific, the pair $[R/v_{\rm Ai},~\rho_{\rm i}/\rho_{\rm e}]$
    reads $[2.94~{\rm secs},~182]$ when $\mu=1$,
    and reads $[1.64~{\rm secs},~88.2]$ when $\mu=100$.
However, the steepness parameter $\mu$ is difficult to constrain, 
    since its possible range is too broad.
This agrees with paper I where we concluded that 
    for spatially unresolved QPPs, the transverse Alfv\'en time 
    is the best constrained, whereas the steepness (the length of the transition layer
    in units of loop radius $l/R$ in that paper)
    corresponds to the other extreme.

\begin{table}[htbp]
\begin{tabular}{c||c|c|c|c|c|c|c|c|c}
  \hline%
 $\mu$ &  1	& 2.5	& 4	& 6.5	& 9	& 12	& 20	& 60	& 100 \\
  \hline%
 $\rho_{\rm i}/\rho_{\rm e}$ 
       & 182.0  & 118.3	& 104.0	& 95.8	& 92.5  & 91.0  & 89.2	& 88.2  & 88.2 \\
  \hline%
 $R/v_{\rm Ai}$ (secs)
       & 2.94   & 2.23	&  2.02	& 1.87	& 1.80	& 1.76	& 1.7	& 1.65	& 1.64	\\
  \hline
\end{tabular}
\caption{
Values of $[\mu,~\rho_{\rm i}/\rho_{\rm e},~R/v_{\rm Ai}]$
   deduced for the QPP event reported in \cite{1973SoPh...32..485M}.}
\label{tab_invcurv}
\end{table}

One may then ask how to make sense of this seismological inversion.
To this end, we may compare our results with what one finds with the DR for a step-function density profile
   (Eq.~18 in paper I).
Noting that the $\mu$-dependence no longer exists in the step-function case, 
    one finds with the measured $P$ and $\tau$ that
    $[R/v_{\rm Ai},~\rho_{\rm i}/\rho_{\rm e}]$ = $[1.62~{\rm secs},~88.2]$.
As expected, this agrees well with what we found for large $\mu$.
However, it differs substantially from the results for small $\mu$.
From this we conclude that, although simple and straightforward,
    the practice for deducing $[R/v_{\rm Ai},~\rho_{\rm i}/\rho_{\rm e}]$
    using the DR for step-function profiles is subject to substantial uncertainty
    if one takes account of the uncertainties in prescribing the transverse density structuring.
In particular, it may substantially underestimate $R/v_{\rm Ai}$.
This uncertainty will be carried over to the deduced values of the Alfv\'en speed
    and consequently the magnetic field strength, provided that one can further
    estimate the loop radius $R$ and internal density $\rho_{\rm i}$.

\subsection{Applications to spatially resolved QPP observations}    
Now move on to the seismological applications of Eqs.~(\ref{eq_Psaus}) and (\ref{eq_tauOPsaus}) to
    spatially resolved QPP events.
In this case the geometrical parameters $L$ and $R$ can be considered known,
    and only the combination of $[v_{\rm Ai},~\mu,~\rho_{\rm i}/\rho_{\rm e}]$ remains to be deduced.
It then follows that if a trapped (leaky) mode is presumably the cause of a QPP event,
    the measured period $P$ ($P$ together with the damping time $\tau$)
    allows a 2D surface (1D curve) to be found in the 3D space
    formed by $[v_{\rm Ai},~\mu,~\rho_{\rm i}/\rho_{\rm e}]$.

Something more definitive can be deduced if a QPP event involves more than just a sausage mode.
Similar to paper I, let us examine the case where a fundamental kink mode exists together
    with a fundamental sausage one, with both experiencing temporal damping.
Let $P_{\rm saus}$ and $\tau_{\rm saus}$ denote the period
    and damping time of the sausage mode, respectively.
Likewise, let $P_{\rm kink}$ ($\tau_{\rm kink}$) denote the
    period (damping time) of the kink mode.
Furthermore, let us assume that wave leakage leads to the apparent damping
    of the sausage mode, whereas resonant absorption is responsible for damping the kink mode.
One finds that $P_{\rm kink}$ and $\tau_{\rm kink}$ can be formally expressed as
\begin{eqnarray}
&&  P_{\rm kink}
    = \displaystyle\frac{L}{v_{\rm Ai}} F_{\rm kink}\left(\frac{L}{R},~\mu,~
    \frac{\rho_{\rm i}}{\rho_{\rm e}}\right),
    \label{eq_Pkink} \\
&& \tau_{\rm kink}
    =\displaystyle\frac{L}{v_{\rm Ai}} H_{\rm kink}\left(\frac{L}{R},~\mu,~
    \frac{\rho_{\rm i}}{\rho_{\rm e}}\right) .
    \label{eq_taukink}
\end{eqnarray}
To establish the functions $F_{\rm kink}$ and $H_{\rm kink}$,
    we adopt the same approach as in \citet{2006ApJ...642..533T}.
A set of linearized resistive MHD equations (Eqs.~1-5 therein)
    is solved for the dimensionless complex angular frequency ($\omega_{\rm kink} L/v_{\rm Ai}$)
    as an eigen-value. 
A uniform resistivity $\bar{\eta}$ is adopted, resulting in a
    magnetic Reynolds number $R_{\rm m} = v_{\rm Ai}R/\bar{\eta}$.
It turns out that $\omega_{\rm kink} L/v_{\rm Ai}$ does not depend on $R_{\rm m}$
    when $R_{\rm m}$ is sufficiently large, and this saturation value
    is taken to be the value that $\omega_{\rm kink} L/v_{\rm Ai}$ attains
    with the input parameters $[L/R,~\mu,~\rho_{\rm i}/\rho_{\rm e}]$
{(see Appendix~\ref{sec_App_resistive} for details).
We note that one can also establish $F_{\rm kink}$ and $H_{\rm kink}$ 
    with the approach developed by \citet{2013ApJ...777..158S},
    where a less computationally costly method based on singular series expansions
    was employed.
}    

With $P_{\rm saus}$, $\tau_{\rm saus}$, $P_{\rm kink}$ and $\tau_{\rm kink}$ measured,
    one finds that the number of equations is more than needed, since now
    there are only three unknowns, $v_{\rm Ai}$, $\mu$, and $\rho_{\rm i}/\rho_{\rm e}$.
In practice, we consider the expression for $P_{\rm kink}$ as the redundant one,
    and use the rest for seismological purposes.
As suggested by paper I, the kink mode period expected from Eq.~(\ref{eq_Pkink})
    with the deduced parameters can be compared with the measured value.
The {difference} between the two allows one to say, for instance,
    whether it is safe to identify the oscillating signals with
    the particular modes.
{In addition, this difference can also serve as an estimate of the errors
    of the deduced loop parameters for a given density prescription.}

\begin{figure}
\centering
\includegraphics[width=0.65\columnwidth]{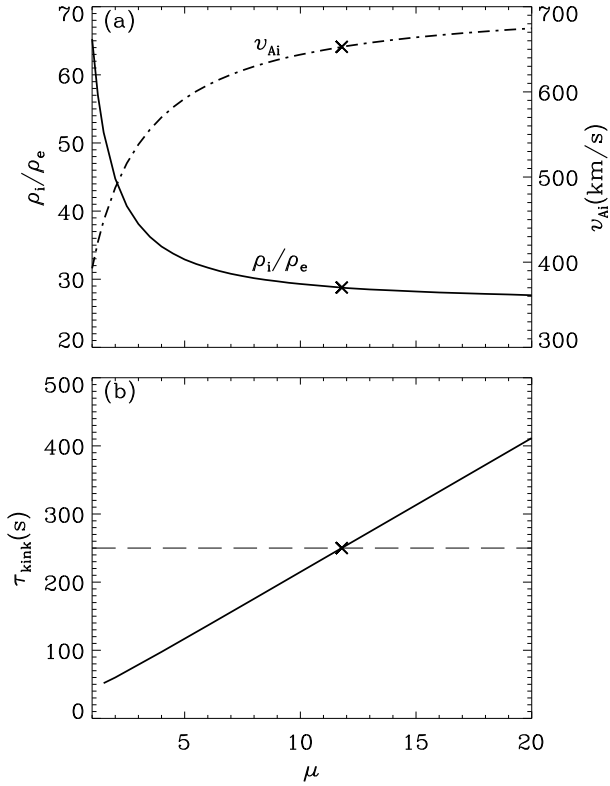}
 \caption{
  Illustration of the scheme for inverting the two-mode QPP event
      reported in~\citet{2015A&A...574A..53K}.
  The curves in (a) are found by requiring that the damping-time-to-period ratio
      $\tau_{\rm saus}/P_{\rm saus}$ and period $P_{\rm saus}$
      for fundamental sausage modes
      to agree with the measured values for a series of given values of $\mu$.
  The solid curve in (b) represents the damping time $\tau_{\rm kink}$
      for fundamental kink modes
      expected with the values $[\mu,~\rho_{\rm i}/\rho_{\rm e},~v_{\rm Ai}]$
      given in (a).
  Its intersection with the horizontal dashed line, representing the measured value
      for $\tau_{\rm kink}$, gives a unique combination
      of $[\mu,~\rho_{\rm i}/\rho_{\rm e},~v_{\rm Ai}]$ labeled by the crosses.
  }
\label{fig_inver}
\end{figure}

While seemingly fortuitous, QPP events involving multiple modes do occur \citep[e.g.,][]{2003A&A...412L...7N,2013SoPh..284..559K,2015A&A...574A..53K}.
For instance, when analyzing the multiple signals in the QPP event on 14 May 2013,
    \citet{2015A&A...574A..53K} identified a fundamental fast kink mode
    with period $P_{\rm kink}=100$~secs and damping time $\tau_{\rm kink}=250$~secs
    together with a fundamental sausage mode
    with $P_{\rm saus}=15$~secs and
    $\tau_{\rm saus}=90$~secs.
In addition, the flare loop hosting the two modes was
    suggested to be of length $L=4\times 10^4$ km
    and radius $R=4\times 10^3$ km, if one takes
    the apparent width of the loop as the loop diameter.
Now the seismological inversion is rather straightforward and involves
    only two steps as illustrated by Fig.~\ref{fig_inver}.
First, with the aid of Eq.~(\ref{eq_tauOPsaus}),
    one readily derives a curve (the solid curve in Fig.~\ref{fig_inver}a)
    in the $[\mu,~\rho_{\rm i}/\rho_{\rm e}]$ plane
    to be compatible with the measured $\tau_{\rm saus}/P_{\rm saus}$. 
The internal Alfv\'{e}n speed $v_{\rm Ai}$ for a given pair of $[\mu,~\rho_{\rm i}/\rho_{\rm e}]$
    is then found with Eq.~(\ref{eq_Psaus}) to agree with the measured $P_{\rm saus}$,
    yielding the dash-dotted curve.
Second, one evaluates the kink mode damping time with Eq.~(\ref{eq_taukink})
    with a series of combinations $[v_{\rm Ai},~\mu,~\rho_{\rm i}/\rho_{\rm e}]$,
    thereby finding the solid curve in Fig.~\ref{fig_inver}b.
The intersection of this solid curve with the horizontal dashed line, representing 
    the measured kink mode damping time ($\tau_{\rm kink}=250$~secs),
    then yields that
    $\mu = 11.8$,
    $\rho_{\rm i}/\rho_{\rm e} = 28.8$,
    and $v_{\rm Ai} = 653~{\rm km}~{\rm s}^{-1}$.
It is worth stressing that  Eq.~(\ref{eq_Pkink}) yields an expected kink mode period of $88$~secs
    with the measured $L$ and $R$ as well as this set of deduced parameters.
This is close to what was measured ($100$~secs), substantiating the interpretation of
    the long-period signal as the fundamental kink mode as done by~\citet{2015A&A...574A..53K}.
{Alternatively, this agreement between the two values also suggests that the errors in this inversion procedure
    are rather moderate.
}    

What are the uncertainties of the derived flare loop parameters?
Evidently, these come entirely from the uncertainties associated
    with the unknown specific form 
    of the transverse density structuring.
To provide an uncertainty measure, 
    we repeat the afore-mentioned inversion process
    for all four different density prescriptions in paper I, where
    we examined only one profile (the sine one) and adopted the 
    TTTB approximation to describe $F_{\rm kink}$ and $H_{\rm kink}$.
Now with the pertinent analytical DRs for sausage modes
    and self-consistent resistive MHD computations for kink modes,
    we find that the density contrast $\rho_{\rm i}/\rho_{\rm e}$
    is constrained to the range from $28.4$ to $31.1$, and the internal Alfv\'en speed
    $v_{\rm Ai}$ lies between $594$ and $658$~${\rm km}~{\rm s}^{-1}$.
Interestingly, for the $\mu$-power profile examined here, the values inferred for
    $\rho_{\rm i}/\rho_{\rm e}$ and $v_{\rm Ai}$ also lie in
    these rather narrow ranges.
On the other hand, the deduced $\mu$ value indicates that the density profile
    across the flare loop in question is rather steep, which also agrees with
    the ratios of the transition layer width to loop radius ($0.167 \le l/R \le 0.284$)
    inferred with the profile prescriptions in paper I. 
From this we conclude that at least for the profiles examined in the present study and paper I,
    the uncertainties of the inferred profile steepness, density contrast,
    and internal Alfv\'en speed are relatively small.

\section{Summary}
\label{sec_con}

A substantial fraction of quasi-periodic pulsations (QPPs) in the lightcurves of solar flares
    is attributed to sausage modes in flare loops.
The present study continues our effort initiated in~\citet[][paper I]{2015ApJ...812...22C}
    to infer flare loop parameters with the measured periods $P$
    and damping times $\tau$ of fundamental standing sausage modes supported therein.
For this purpose we extended the analysis in paper I to sausage waves 
    in nonuniform, straight, coronal cylinders with arbitrary transverse density profiles
    comprising a nonuniform inner portion and a uniform external medium.
Working in the framework of ideal, cold, MHD, 
    we derived an analytical dispersion relation (DR, Eq.~\ref{eq_DR})
    and focused on density profiles of a $\mu$-power form (Eq.~\ref{eq_def_fr}).
The dispersive properties of fundamental, standing modes were examined, together
    their potential for inferring flare loop parameters.

We found that $P$ and $\tau$ in units of the transverse Alfv\'en time $R/v_{\rm Ai}$
    depend only on the density contrast $\rho_{\rm i}/\rho_{\rm e}$,
    length-to-radius ratio $L/R$ of coronal cylinders,
    and the profile steepness $\mu$.
For all profiles examined in both this study and paper I,
    when the rest of the parameters are fixed,
    $P$ ($\tau$) in units of $R/v_{\rm Ai}$ increases (decreases) with increasing $L/R$ and 
    tends to some saturation value when $L/R$ is sufficiently large.
For spatially unresolved QPPs, we showed that 
    one can at most deduce
    a curve in the 3-dimensional space formed by $R/v_{\rm Ai}$,
    $\rho_{\rm i}/\rho_{\rm e}$, and $\mu$.
This happens when one can assume that $L/R \gg 1$ beforehand.     
Applying this inversion procedure to the event reported by \citet{1973SoPh...32..485M},
    we found that $R/v_{\rm Ai}$ is the best constrained, whereas the steepness parameter
    is the least constrained.
For spatially resolved QPPs, we showed that while geometric parameters of flare loops
    are available, the inversion problem remains under-determined.
However, when an additional mode co-exists with the fundamental sausage mode,
    the full information on the unknowns, $[v_{\rm Ai},~\mu,~\rho_{\rm i}/\rho_{\rm e}]$,
    can be inferred.
In fact, in this case the inversion problem may become over-determined.
Applying this idea to a recent QPP event where temporally decaying kink and sausage modes
    were identified, we found that $v_{\rm Ai}$, $\rho_{\rm i}/\rho_{\rm e}$,
    and the profile steepness can be constrained to rather narrow ranges.
    
The discussions on the limitations to our inversion procedures as presented in paper I also apply here
   and will not be repeated.
Instead, let us {stress the}
   great potential of using multi-mode QPP measurements
   to determine flare loop parameters rather precisely, the internal Alfv\'en speed in particular.
To this end, not only modes of distinct nature (e.g., a fundamental kink mode co-existing with
   a sausage one) are useful, modes of the same nature but with different longitudinal node numbers
   can also do the job.
While fundamental kink modes and their harmonics
   have been seismologically exploited \citep[see e.g., the review by][]{2009SSRv..149....3A},
   serious studies using sausage modes
   need to be conducted.  
   
{Before closing, we note that Bayesian techniques have been successfully applied to
   the inference of density structuring transverse to coronal loops hosting
   resonantly damping kink modes \citep{2013A&A...554A...7A,2013ApJ...769L..34A,2015ApJ...811..104A}.
With such techniques, the errors in the measurements of kink mode periods and damping times
   can be properly propagated,
   and the plausibility of a density profile prescription can be assessed.
When no particular prescription is favored, approaches like model-averaging
   can be employed to yield an evidence-averaged inference.
While so far the applications of such techniques have been primarily focused on kink modes,
   similar ideas are expected to be equally applicable to sausage modes.
For this purpose, the DRs derived here and in paper I should be useful.
}

\begin{acks}
    {We thank the referee for his/her constructive comments, which helped improve this manuscript substantially.}
    This research is supported by the 973 program 2012CB825601, National Natural Science Foundation of China
    (41174154, 41274176, 41274178, and 41474149),
    the Provincial Natural Science Foundation of Shandong via Grant JQ201212,
    and also by a special fund of Key Laboratory of Chinese Academy of Sciences.
\end{acks}


\appendix
\section{Fast sausage modes in nonuniform cylinders: a time-dependent approach}
\label{sec_App_tdependent}

\begin{figure}
\centering
\includegraphics[width=0.65\columnwidth]{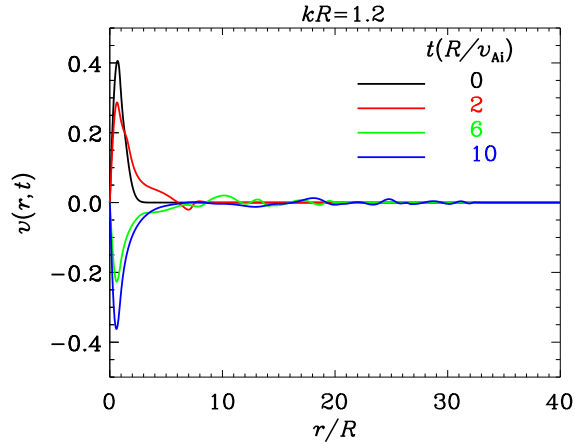}
 \caption{
 Spatial distribution of the transverse velocity perturbation at a number of different times
     for $kR = 1.2$.
 The initial perturbation is described by Eq.~(\ref{eq_tdep_IC}).     
 Here the density ratio $\rho_{\rm i}/\rho_{\rm e}=10$
     and the steepness parameter $\mu = 3$.
  }
\label{fig_tdpen_trap_spatial}
\end{figure}

This section provides a detailed examination of sausage modes
    from an initial-value-problem perspective.
We note that similar studies were carried out for step-function density profiles
    by \citet{2007SoPh..246..231T}, and for continuous profiles by 
    \citet{2012ApJ...761..134N,2015SoPh..290.2231C}.
To start, it is straightforward to derive an equation governing the transverse velocity perturbation $\delta v_r(r, z, t)$ 
    from linearized, time-dependent, cold MHD equations.
Formally expressing $\delta v_r(r, z, t)$ as $v(r,t) \sin(kz)$, 
    one finds that $v(r,t)$ is governed by~\citep[e.g.,][]{2015SoPh..290.2231C} 
\begin{equation}\label{eq_xi(r,t)}
   \displaystyle\frac{\partial^2 v(r,t)}{\partial t^2}
 = v^2_{\rm A}(r)\left[\displaystyle\frac{\partial^2}{\partial r^2}
 +\displaystyle\frac{1}{r}\displaystyle\frac{\partial}{\partial r}
 -\left(\displaystyle\frac{1}{r^2}+k^2\right)\right]
       v(r,t)~.
\end{equation}
With a $\rho(r)$ profile given by Eqs.~(\ref{eq_density}) and (\ref{eq_def_fr}),
    one readily evaluates the profile
    for the Alfv\'en speed $v_{\rm A}(r) = B/\sqrt{4\pi \rho(r)}$.
Equation~(\ref{eq_xi(r,t)}) can then be readily solved
    when supplemented with appropriate initial and boundary conditions.
For this purpose, we developed a simple finite-difference code second-order accurate in both space and time,
    and solve Eq.~(\ref{eq_xi(r,t)}) on a uniform grid spanning $[0, r_{\rm outer}]$ with a spacing $\Delta r = 0.02~R$
    and $r_{\rm outer} = 1000~R$.
A uniform time-step $\Delta t = 0.8 \Delta r/v_{\rm Ae}$ is adopted to ensure numerical stability in view of 
    the Courant condition.
We have made sure that further refining the grid leads to no discernible difference.
Furthermore, the outer boundary $r_{\rm outer}$ is placed sufficiently far from the cylinder such that
    the signals to be analyzed are not contaminated by the perturbations reflected off the outer boundary.
Pertinent to sausage modes, we require that $v(r=0, t) = 0$.
In addition, $v(r=r_{\rm outer}, t)$ is specified to be zero for simplicity.
Throughout this section, we examine a density ratio $\rho_{\rm i}/\rho_{\rm e}$ of $10$,
    and a steepness parameter $\mu$ of $3$. 
Moreover, for all computations we adopt the same initial condition (IC)
\begin{eqnarray}
\label{eq_tdep_IC}
   v(r, t=0) = \frac{r}{R}\exp\left[-\left( \frac{r}{R} \right)^2\right]~,~~~~
   \frac{\partial}{\partial t} v(r, t=0) = 0
\end{eqnarray}
   which is chosen not to be too localized to avoid exciting higher order modes.

Figure~\ref{fig_tdpen_trap_spatial} presents the spatial distribution of $v(r, t)$ for $kR = 1.2$
    at a number of $t$ as labeled.
One sees that as time progresses, some ripples propagate outward with the external 
    Alfv\'en speed ($v_{\rm Ae}=\sqrt{\rho_{\rm i}/\rho_{\rm e}} v_{\rm Ai} = \sqrt{10}v_{\rm Ai}$).
The amplitudes of these ripples are rather insignificant and decrease with time, meaning that 
    little energy is transmitted into the external medium 
    for the adopted IC even though it 
    is not an exact eigen-function.
The majority of the energy is trapped in the cylinder, a signature of trapped modes. 

\begin{figure}
\centering
\includegraphics[width=0.65\columnwidth]{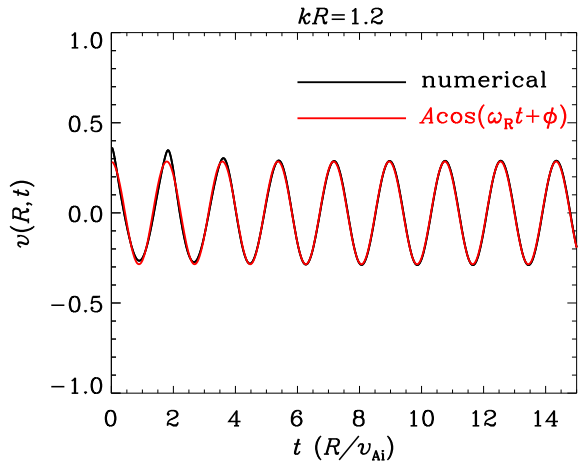}
 \caption{
 Temporal evolution of the transverse velocity perturbation $v(r=R,t)$ 
     for $kR = 1.2$.
 The initial perturbation is described by Eq.~(\ref{eq_tdep_IC}).     
 In addition to the numerical result from this time-dependent computation (the black curve),
     a fit to this curve in the form $A\cos(\omega_{\rm R}t+\phi)$ is
     given by the red line for comparison.
 Here the density ratio $\rho_{\rm i}/\rho_{\rm e}=10$
     and the steepness parameter $\mu = 3$.
  }
\label{fig_tdpen_trap_temporal}
\end{figure}

That this computation pertains to the trapped regime is better shown by 
    Fig.~\ref{fig_tdpen_trap_temporal}, where the temporal evolution of $v(R, t)$ is displayed.
In addition to the numerical results (the black curve), a fit in the form $A\cos(\omega_{\rm R} t + \phi)$
    is given by the red line.
This fitting procedure yields that $\omega_{\rm R} = 3.51 v_{\rm Ai}/R$, in exact
    agreement with the value  
    found from solving the dispersion relation (Eq.~\ref{eq_DR}).
One sees that the black curve can be hardly told apart from the red one
    when $t\gtrsim 4 R/v_{\rm Ai}$,
    meaning that the signal at this location rapidly evolves
    into a trapped eigenmode.
    
\begin{figure}
\centering
\includegraphics[width=0.65\columnwidth]{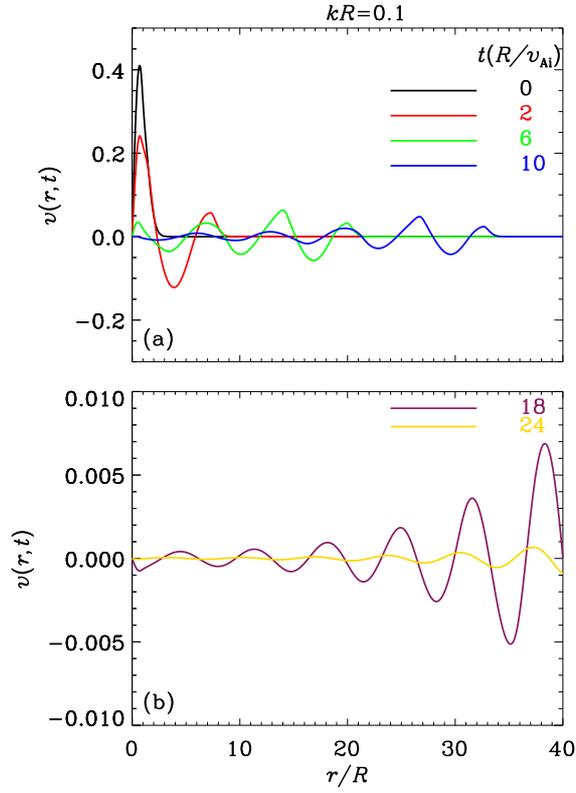}
 \caption{
 Similar to Fig.~\ref{fig_tdpen_trap_spatial} but for $kR = 0.1$. 
 Note that the spatial distributions for $t=18$ and $24~R/v_{\rm Ai}$
     are plotted in a separate panel.
  }
\label{fig_tdpen_leaky_spatial}
\end{figure}

What happens for a small $kR$?
This is examined in Fig.~\ref{fig_tdpen_leaky_spatial} where the spatial dependence of $v(r, t)$ 
    for $kR = 0.1$ is presented.
In response to the initial perturbation, some ripples
    are also seen to propagate away from the cylinder.
However, in this case the signal close to the cylinder axis ($r=0$) decays  
    so rapidly that a different scale has to be used
    to plot $v(r, t)$ at large times (Fig.~\ref{fig_tdpen_leaky_spatial}b).
From Fig.~\ref{fig_tdpen_leaky_spatial}b one also sees that at a given time,
    the amplitude of the perturbations in the external medium 
    tends to increase with distance first before decreasing towards
    the front (also see the blue curve in Fig.~\ref{fig_tdpen_leaky_spatial}a).
This is a signature of leaky eigen-functions \citep[e.g.,][]{1986SoPh..103..277C,2007SoPh..246..231T}.

\begin{figure}
\centering
\includegraphics[width=0.65\columnwidth]{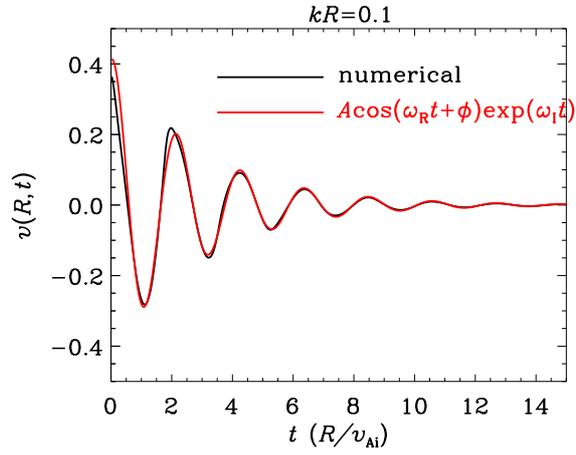}
 \caption{
 Similar to Fig.~\ref{fig_tdpen_trap_temporal} but for $kR = 0.1$. 
 Note that the fit (the red curve) to the time-dependent solution (black)
    is in the form $A\cos(\omega_{\rm R}t+\phi)\exp(\omega_{\rm I}t)$.
  }
\label{fig_tdpen_leaky_temporal}
\end{figure}

Figure~\ref{fig_tdpen_leaky_temporal} presents the temporal evolution of $v(R, t)$ (the black curve)
    together with a fitting in the form 
    $A\cos(\omega_{\rm R} t + \phi)\exp(\omega_{\rm I} t)$ (red).
From this fitting we find that 
    $[\omega_{\rm R},\omega_{\rm I}]=[2.98, -0.34] v_{\rm Ai}/R$, coinciding with
    the eigenmode computation for the adopted $kR$.
One sees that the black and red curves agree closely with each other
    for $t\gtrsim 2.5 R/v_{\rm Ai}$, substantiating the interpretation that
    the signal settles to a leaky eigenmode.

\section{Fast kink modes in nonuniform cylinders: a resistive, linear MHD computation}
\label{sec_App_resistive}

This section provides some details for the resistive, linear MHD computations that we
    employ to establish the functions $F_{\rm kink}$ and $H_{\rm kink}$ contained in Eqs.~(\ref{eq_Pkink})
    and (\ref{eq_taukink}).
Such a description seems informative even though our approach is identical to the one adopted by
    \citet[][hereafter TOB06]{2006ApJ...642..533T},
    since the density profile given by Eq.~(\ref{eq_density}) has not been explored 
    for resonantly damped kink modes.
Now that the approach has been detailed in section 3.1 in TOB06, it suffices to note here that 
    we are looking for a dimensionless complex-valued angular frequency $\omega_{\rm kink} L/v_{\rm Ai}$
    for a set of dimensionless parameters $[L/R, \mu, \rho_{\rm i}/\rho_{\rm e}]$.
The magnetic Reynolds number $R_{\rm m} = v_{\rm Ai} R/\bar{\eta}$ is also relevant, where 
    $\bar{\eta}$ is the resistivity and assumed to be uniform.

\begin{figure}
\centering
\includegraphics[width=0.65\columnwidth]{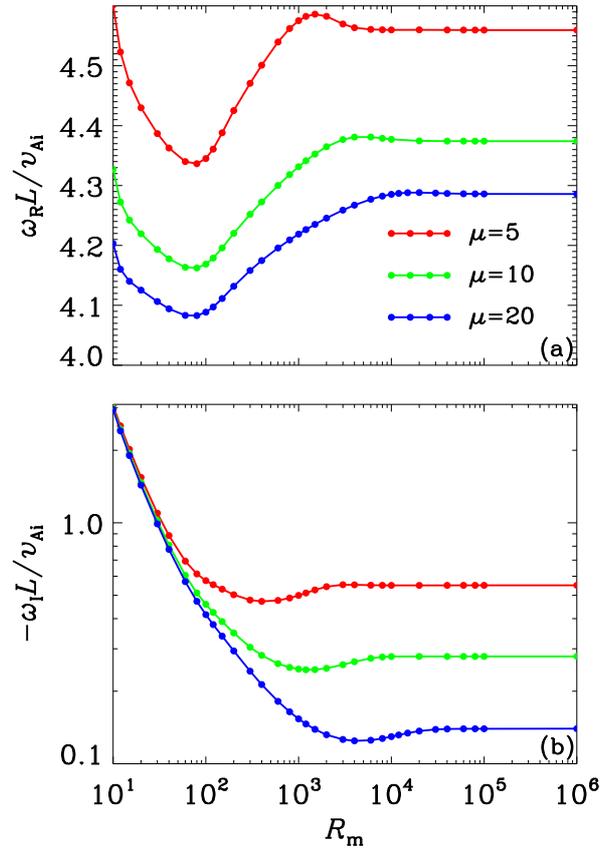}
 \caption{
 Dependence on the magnetic Reynolds number $R_{\rm m}$ of (a) the real
    and (b) imaginary parts of the dimensionless eigen-frequency
    for kink modes in cylinders with
    a transverse density profile given by Eqs.~(\ref{eq_density}) and (\ref{eq_def_fr}).
 A number of different steepness parameters $\mu$ are examined as labeled.   
 Here the dimensionless longitudinal wavenumber $kR = 0.1\pi$, 
    and the density ratio $\rho_{\rm i}/\rho_{\rm e}$ is fixed at $20$.
 }
\label{fig_resis_RmDep}
\end{figure}

\begin{figure}
\centering
\includegraphics[width=0.65\columnwidth]{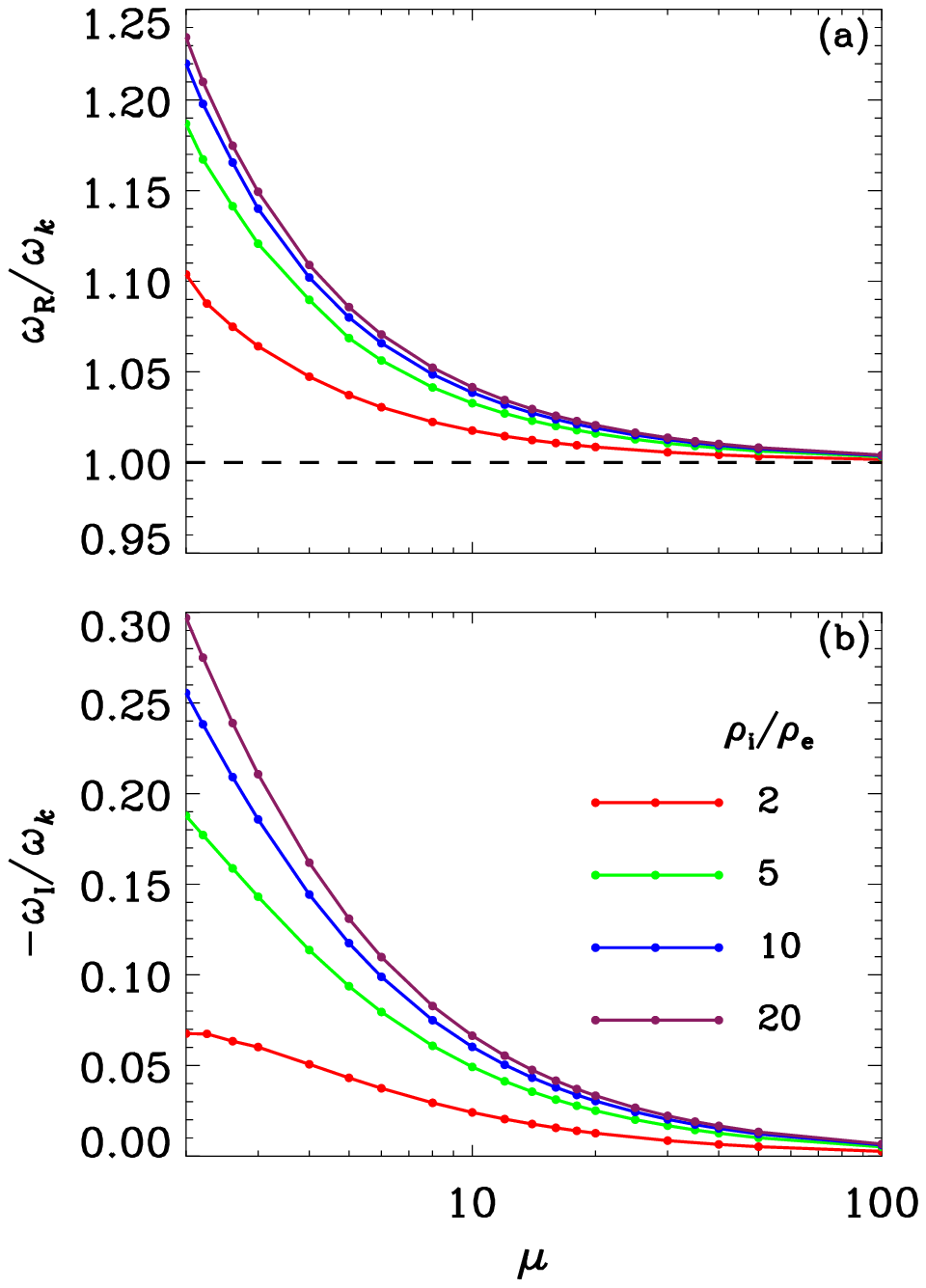}
 \caption{
 Dependence on the steepness parameter $\mu$ of (a) the real
    and (b) imaginary parts of the eigen-frequencies
    for resonantly damped kink modes in cylinders with
    a transverse density profile given by Eqs.~(\ref{eq_density}) and (\ref{eq_def_fr}).
 A number of density ratios $\rho_{\rm i}/\rho_{\rm e}$ are examined as labeled.   
 Here the dimensionless longitudinal wavenumber $kR = 0.1\pi$.
 The eigen-frequencies are found with linear resistive MHD computations at sufficiently large
    magnetic Reynolds numbers.
 They are normalized to $\omega_k$, the value attained for density profiles of a step-function form.
 See text for details.
 }
\label{fig_resis_steepnessDep}
\end{figure}

Figure~\ref{fig_resis_RmDep} presents the $R_{\rm m}$ dependence of the real ($\omega_{\rm R}$) and imaginary ($\omega_{\rm I}$)
    parts of the dimensionless angular frequency 
    for $kR = 0.1\pi$ and $\rho_{\rm i}/\rho_{\rm e} = 20$.
A number of $\mu$ values are examined as given by the curves in different colors.
From Fig.~\ref{fig_resis_RmDep}b one sees that the curves show a significant $R_{\rm m}$ dependence
    only for relatively small $R_{\rm m}$.
As discussed in TOB06, this is attributable to the competition between resistivity and resonant absorption
    in damping the kink modes.
For large (small) $R_{\rm m}$, resonant absorption (resistivity) plays a more important role
    and consequently the damping rate $|\omega_{\rm I}|$ is insensitive (sensitive) to $R_{\rm m}$.
Interestingly, in agreement with Fig.~2 of TOB06, $|\omega_{\rm I}|$ is not sensitive to the density profile steepness
    when $R_{\rm m}$ is small.
On the other hand, $|\omega_{\rm I}|$ rapidly settles to some 
    asymptotic value when $R_{\rm m}$ exceeds some critical value.
Similar to TOB06, with increasing profile steepness
    this critical $R_{\rm m}$ increases, whereas the asymptotic $|\omega_{\rm I}|$ decreases.
Examining Fig.~\ref{fig_resis_RmDep}a, one finds that $\omega_{\rm R}$ is different for different $\mu$ values even
    at small $R_{\rm m}$.
Despite this, what is important for the present purpose is that neither $\omega_{\rm R}$
    nor $\omega_{\rm I}$ depends on $R_{\rm m}$ when $R_{\rm m}$ is sufficiently large. 
And their asymptotic values are taken to be the eigen-frequencies of kink modes
    whose damping is solely due to resonant absorption.

How do these saturation values depend on the steepness parameter $\mu$ when the rest of the parameters are fixed?
This is examined in Fig.~\ref{fig_resis_steepnessDep} where a series of computations are conducted for a number of
    density ratios $\rho_{\rm i}/\rho_{\rm e}$ as labeled.
The dimensionless longitudinal wavenumber $kR$ is also taken to be $0.1\pi$.    
For presentation purposes, both $\omega_{\rm R}$ and $\omega_{\rm I}$     
    are normalized to the kink frequency $\omega_k$, which is attained for density profiles
    of a step-function form. 
It turns out that the correction to $\omega_k$ due to dispersion is not negligible at the chosen $kR$, 
    meaning that $\omega_k$ needs to be computed by solving the relevant dispersion relation 
    \citep[e.g.,][Eq.~8b]{1983SoPh...88..179E}.
The computed value is a few percent different from its thin-tube counterpart, namely
    $\sqrt{2} k v_{\rm Ai}/\sqrt{1+\rho_{\rm e}/\rho_{\rm i}}$
    \citep[e.g., Eq.~40 in][hereafter S13]{2013ApJ...777..158S}.    
Similar to Fig.~1 in S13, our Fig.~\ref{fig_resis_steepnessDep} shows that $\omega_{\rm R}$
    and $|\omega_{\rm I}|$ tend to decrease with increasing density profile steepness, 
    approaching the step-function values when $\mu$ is sufficiently large.
In addition, $|\omega_{\rm I}|/\omega_k$ at a fixed steepness parameter increases 
    with increasing $\rho_{\rm i}/\rho_{\rm e}$.    
However, while here $\omega_{\rm R}/\omega_k$ tends to increase with $\rho_{\rm i}/\rho_{\rm e}$ regardless of $\mu$,
    it does not show a monotonical dependence on $\rho_{\rm i}/\rho_{\rm e}$ in Fig.~1 of S13
    when the steepness parameter is fixed.
This difference signifies the importance of the detailed description of the density profile
    in determining the properties of resonantly damped kink modes:
    while a $\mu$-power profile is examined here, S13 explored a sine profile (Eq.~63 therein).

\section{Examining sausage modes in nonuniform cylinders with series-expansion-based methods}
\label{sec_App_compareSeries}

So far two series-expansion-based methods have been available to derive 
    explicit expressions for 
    the sausage perturbations in the nonuniform portion of the density distribution.
One (approach I, S13) is based on singular expansions as a byproduct of a comprehensive examination
    of resonantly damped kink modes,
    whereas the other (approach II, paper I) is based on regular expansions.
This section provides a rather detailed comparison between the two.

To facilitate this comparison, let us focus on density profiles considered by both studies,
    where a transition layer (TL) connects a uniform cord (with density $\rho_{\rm i}$)
    and a uniform external medium (with density $\rho_{\rm e}$).
The TL is of width $l$ and centers around $r=R$.
Approach I solves the perturbation equation (Eq.~4 in S13) by conducting an expansion about 
     the Alfv\'en resonance $r_{\rm A}$ where $\omega_{\rm R} = k v_{\rm A}$.
For the density profiles in question, $r_{\rm A}$ is located in the TL.     
Valid for arbitrary azimuthal wavenumbers $m$, the analysis in S13 yields that for sausage modes ($m=0$),
     the series solutions are regular even though $r_{\rm A}$ is a regular singular point. 
Physically, this means that sausage modes do not resonantly couple to the Alfv\'en continuum.
Approach II capitalizes on this fact and solves the perturbation equation (Eq.~6 in paper I) by performing
     a regular series expansion about $r=R$.
In this aspect, approach II is equivalent to I and both should yield identical solutions, provided that
     a point exists in the TL such that $\omega_{\rm R} = k v_{\rm A}$.
Let $r_{\rm A}$ denote this point for brevity,      
     although it is not a resonance for sausage modes. 

Before proceeding, we note that the perturbations in the external medium
    were required to be evanescent by S13, since leaky modes
    were not of interest therein.
Consequently, the Fourier amplitude of the Eulerian perturbation of total pressure
    was expressed with $K_0(k_{\perp, {\rm e}} r)$, the modified Bessel function of the first kind
    (Eq.~10 in S13).
To account for leaky sausage modes, this needs to be replaced with $H_0^{(1)}(\mu_{\rm e}r)$
    \citep[e.g.,][]{1986SoPh..103..277C}.
Here $\mu_{\rm e}$ is defined by Eq.~(\ref{eq_def_mue}), and by definition
    $\mu_{\rm e}^2 = -k_{\perp, {\rm e}}^2$.     
With the notations in S13, the DR therein then reads
\begin{eqnarray}
&& 
\displaystyle
  \frac{
        \displaystyle
	\frac{-\mu_{\rm e}}{\rho_{\rm e}\left(\omega^2-k^2 v_{\rm A,e}^2\right)}
        \displaystyle
	\frac{H^{(1)}_1\left[\mu_{\rm e}\left(R+l/2\right)\right]}
	    {H^{(1)}_0\left[\mu_{\rm e}\left(R+l/2\right)\right]}
       {\cal G}_{\rm e}-\Xi_{\rm e}
      }
      {
        \displaystyle
        \frac{-\mu_{\rm e}}{\rho_{\rm e}\left(\omega^2-k^2 v_{\rm A,e}^2\right)}
        \displaystyle
	\frac{H^{(1)}_1\left[\mu_{\rm e}\left(R+l/2\right)\right]}
	    {H^{(1)}_0\left[\mu_{\rm e}\left(R+l/2\right)\right]}
        {\cal F}_{\rm e}-\Gamma_{\rm e}
      }             \nonumber \\
&&
-\displaystyle
  \frac{
        \displaystyle
	\frac{-k_{\perp, \rm i}}{\rho_{\rm i}\left(\omega^2-k^2 v_{\rm A,i}^2\right)}
        \displaystyle
	\frac{J_1\left[k_{\perp, \rm i}\left(R-l/2\right)\right]}
	     {J_0\left[k_{\perp, \rm i}\left(R-l/2\right)\right]}
       {\cal G}_{\rm i}-\Xi_{\rm i}
      }
      {
        \displaystyle
	\frac{-k_{\perp, \rm i}}{\rho_{\rm i}\left(\omega^2-k^2 v_{\rm A,i}^2\right)}
        \displaystyle
	\frac{J_1\left[k_{\perp, \rm i}\left(R-l/2\right)\right]}
	     {J_0\left[k_{\perp, \rm i}\left(R-l/2\right)\right]}
        {\cal F}_{\rm i}-\Gamma_{\rm i}
      }             \nonumber \\
&&     =0~~. \label{eq_DR_S13gen}      
\end{eqnarray}
For non-leaky waves, this recovers Eq.~(27) in S13, given that
\begin{eqnarray*}
     \frac
     {\mu_{\rm e} H_1^{(1)}\left[\mu_{\rm e}\left(R+l/2\right)\right]}
     {H_0^{(1)}\left[\mu_{\rm e}\left(R+l/2\right)\right]}
  =  \frac
     {k_{\perp,\rm{e}} K_1\left[k_{\perp,\rm{e}}\left(R+l/2\right)\right]}
     {K_0\left[k_{\perp,\rm{e}}\left(R+l/2\right)\right]}~~,
\end{eqnarray*}
    where we have used the relation \citep[p. 375 in][]{1970hmfw.book.....A}
\begin{eqnarray*}
   K_m(w) = \frac{\pi}{2}i^{m+1}H_m^{(1)}(iw)~~~(-\pi < \arg w \le \frac{\pi}{2},~m=0, 1, \cdots)~.
\end{eqnarray*}
For illustration purposes, in what follows we examine a linear profile for the density distribution
    in the TL (Eq.~4 in paper I), and assume that $\rho_{\rm i}/\rho_{\rm e}=100$.
For approach I, we solve Eq.~(\ref{eq_DR_S13gen}) instead of Eq.~(27) in S13,
    and for approach II we solve our Eq.~(17) in paper I.

\begin{figure}
\centering
\includegraphics[width=0.65\columnwidth]{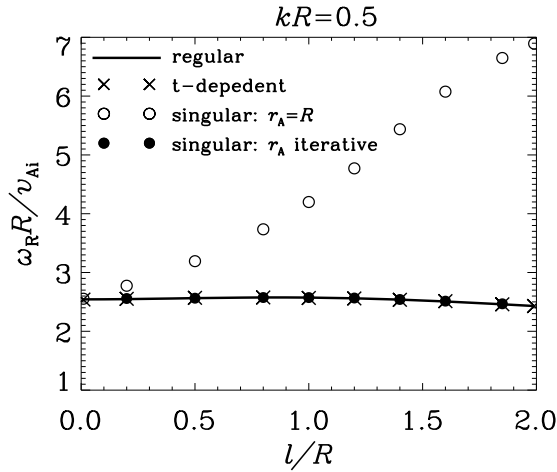}
 \caption{
 Comparison of two series-expansion-based methods for computing eigen-frequencies of sausage modes.
 To facilitate this comparison, the density distribution is   
      different from the one described by Eq.~(\ref{eq_density}).
 Instead, the profile labeled ``linear'' in Eq.~(4) of paper I is adopted.
 Here the dimensionless longitudinal wavenumber $kR=0.5$, pertinent to the trapped regime ($\omega_{\rm I}=0$)
      for the chosen density ratio $\rho_{\rm i}/\rho_{\rm e}$ of $100$.
 The real part of the eigen-frequency $\omega_{\rm R}$ is displayed as a function of $l/R$, the density 
      lengthscale in units of loop radius.
 The black curve represents the results found with the method based on regular expansions,
      and the crosses represent those obtained with analyzing the perturbation signals
      in the corresponding time-dependent computations.
 Two treatments are adopted for the method based on singular expansions.
 In one treatment the location of the nominal Alfv\'en resonance $r_{\rm A}$
      is supposed to be $R$ (the open dots), while in the other it is found iteratively (filled).
 See text for details.
      }
\label{fig_S13_trapped}
\end{figure}

\begin{figure}
\centering
\includegraphics[width=0.65\columnwidth]{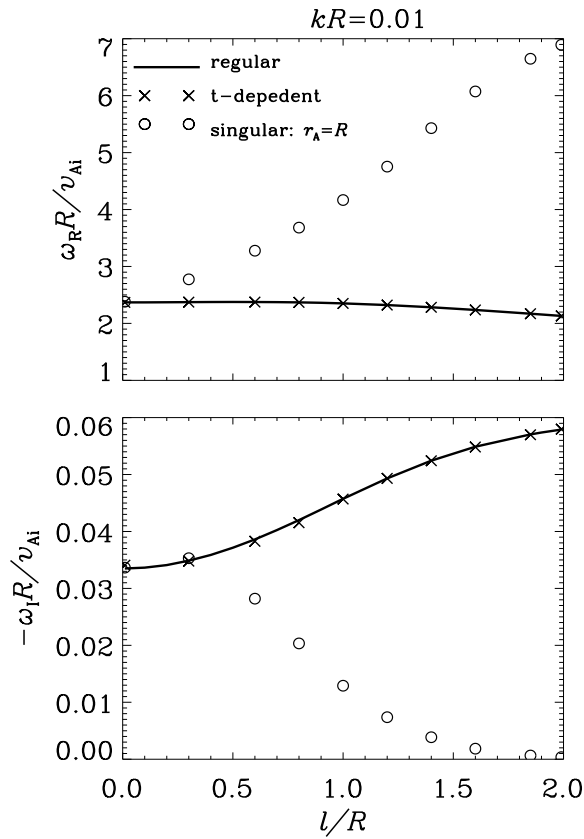}
 \caption{
 Similar to Fig.~\ref{fig_S13_trapped} but for $kR = 0.01$ pertinent to leaky modes, for which 
      $\omega_{\rm I}$ is non-zero and displayed in a separate panel. 
 See text for details.
  }
\label{fig_S13_leaky}
\end{figure}

Figure~\ref{fig_S13_trapped} compares the eigen-frequencies found for a series of $l/R$
     from the two approaches when $kR$ equals $0.5$.
This $kR$ falls into the trapped regime since the imaginary parts ($\omega_{\rm I}$)
     of the eigen-frequencies are zero, and hence only the real parts ($\omega_{\rm R}$) are shown.
The solid line labeled ``regular'' represents the solutions from approach II.
When adopting approach I, 
     we examine two different treatments for the location of $r_{\rm A}$:
     in one we simply take $r_{\rm A}$ to be $R$ (the open dots),
     whereas in the other we solve Eq.~(\ref{eq_DR_S13gen}) iteratively to simultaneous derive $\omega_{\rm R}$
     and $r_{\rm A}$ (the filled dots).
Note that a value for $r_{\rm A}$ needs to be specified before solving this DR.
However, the Alfv\'en speed at this guessed location usually does not 
     equal to $\omega_{\rm R}/k$ thus found.
Hence this latter iterative treatment.
One sees that the filled dots fall on the solid curve, and both are in exact agreement with 
      the crosses representing the values found from fitting the signals $v(R, t)$
      in the corresponding time-dependent computations (see Appendix~\ref{sec_App_tdependent}).
This means that both approaches yield correct results, and approach II can be seen as a specific case
      of the more general analysis presented in S13.
Nonetheless, the advantage of approach II is that there is no need to find $r_{\rm A}$, which is necessary for 
      approach I, given that simply assuming $r_{\rm A} = R$ beforehand
      can yield considerably different results (see the open dots).

Some considerable difference arises when leaky sausage modes are examined.
Let us consider $kR=0.01$, for which the real ($\omega_{\rm R}$)
      and imaginary ($\omega_{\rm I}$) parts of the complex-valued eigen-frequencies
      are presented in Fig.~\ref{fig_S13_leaky}. 
The results from approach II are given by the solid line,
      and are found to agree well with the values found from
      analyzing the time-dependent results (the crosses).
However, they differ substantially from the results found with approach I where
      we assume that $r_{\rm A}=R$.
In this case the iterative treatment does not work because 
      no point in the TL corresponds to a $v_{\rm A}$ that equals $\omega_{\rm R}/k$.
The reason is that in the leaky regime, the apparent phase speed $\omega_{\rm R}/k$
      consistently exceeds $v_{\rm Ae}$, which in turn always exceeds 
      the Alfv\'en speeds in the TL.

Despite the afore-mentioned discussions, we stress that the mathematical approach presented in S13
      is sufficiently general to treat modes with arbitrary azimuthal wavenumber $m$, trapped sausage
      modes ($m=0$) included.
A singular series expansion is necessary for treating all modes with $m \ne 0$.
      

\bibliographystyle{spr-mp-sola}
\bibliography{rev1}

\IfFileExists{\jobname.bbl}{} {\typeout{}
\typeout{****************************************************}
\typeout{****************************************************}
\typeout{** Please run "bibtex \jobname" to obtain} \typeout{**
the bibliography and then re-run LaTeX} \typeout{** twice to fix
the references !}
\typeout{****************************************************}
\typeout{****************************************************}
\typeout{}}

\end{article}

\end{document}